\documentclass[twocolumn,showpacs,preprintnumbers,amsmath,amssymb,epsf]{revtex4-1}

\usepackage{graphicx,epsfig}
\usepackage{dcolumn}
\usepackage{bm}
\newcommand{\be}{\begin{equation}}
\newcommand{\ee}{\end{equation}}
\newcommand{\bse}{\begin{subequations}}
\newcommand{\ese}{\end{subequations}}
\newcommand{\bary}{\begin{eqnarray}}
\newcommand{\eary}{\end{eqnarray}}

\begin{document}


\title{Effect of Resonant Neutrino Oscillation on TeV Neutrino Flavor
  Ratio \\from Choked GRBs}
\author{Sarira Sahu$^{*}$ and Bing Zhang$^{\dagger}$}

\affiliation{$^*$Instituto de Ciencias Nucleares,
Universidad Nacional Aut\'onoma de M\'exico, 
Circuito Exterior, C.U., A. Postal 70-543, 04510 Mexico DF, Mexico\\
$^{\dagger}$Department of Physics and Astronomy,
University of Nevada, Las Vegas, NV 89154, USA
}


\begin{abstract}

In the collapsar scenario of the long duration Gamma-Ray Bursts 
(GRBs), multi-TeV neutrino emission is predicted as the jet makes its way
through the stellar envelope. Such a neutrino signal
is also expected for more general ``failed'' GRBs in which a
putative jet is ``choked'' by a heavy envelope.
If the $\nu_e \rightarrow \nu_\mu$ neutrino oscillation parameters 
are in the atmospheric neutrino
oscillation range, we show that the resonant oscillation of
$\nu_e\leftrightarrow\nu_{\mu,\tau}$ can take place within the inner
high density region of the choked jet progenitor with a heavy envelope,
altering the $\nu$ flavor ratio on its
surface to $\Phi^s_{\nu_e}:\Phi^s_{\nu_\mu}:\Phi^s_{\nu_\tau}=5:11:2$.
Considering vacuum oscillation of these neutrinos on their way to
Earth, the final flavor ratio detected on Earth is further modified 
to either $1:1.095:1.095$ for the
large mixing angle solution to the solar neutrino data, or 1:1.3:1.3 for
maximal mixing among the muon and tau neutrinos in vacuum.

\end{abstract}

\maketitle

\section{Introduction}

Long duration GRBs (LGRBs) are believed to be associated with deaths 
of massive stars \cite{Woosley:1993wj}. The evidence in support of 
such an origin includes associations
of several LGRBs with Type Ic supernovae and the prevalence of
star forming dwarf host galaxies associated with LGRBs
\cite{Woosley:2006fn}. 
Observationally, only a small fraction ($\leq 10^{-3}$) of core
collapse SNe are associated with GRBs \cite{Berger:2003kg}.
They correspond to those jets 
that break through the stellar 
envelope and reach a highly relativistic speed 
(Lorentz factor $\Gamma \geq 100$). Internal shocks are formed
in the optically thin regions, and gamma-rays are produced by synchrotron 
radiation and/or inverse-Compton scattering of Fermi accelerated
electrons in these shocks. 

On the other hand, it
is feasible to envision that a much larger fraction of core
collapses may also launch a mildly relativistic jet from the 
central engine, but the jet never makes its way out from the
envelope, due to either a smaller energy budget or a more extended,
massive stellar envelope than in a GRB progenitor. In any case, 
both the successful and these ``choked'' jets can accelerate 
protons to energy $\geq 10^5$ GeV from the internal shocks 
well inside the stellar envelope. The interaction between these
protons and the $\sim 1$ keV thermal X-ray photons emitted by the
hot cocoon surrounding the jet would generate 
multi-TeV neutrinos through photopion production 
\cite{Meszaros:2001ms}. For an individual GRB at redshift $z\sim 1$,
the predicted upward going muon event number is 0.1-10 in a km$^3$ 
detector \cite{Meszaros:2001ms}.
There can also be neutrino production due to $pp$ and $pn$ collisions
involving relativistic protons from the buried jet and the thermal 
nucleons from the jet and the surrounding, which can produce
more abundant neutrinos for the presupernova stars with a 
heavy envelope \cite{Razzaque:2004yv}. 
The detection of low luminosity (LL) GRBs (such as GRB 980425 and GRB
060218 \cite{Campana:2006qe}) suggests that the event rate of
gamma-ray dim core collapses is much higher than those associated
with high luminosity GRBs \cite{Liang:2007}. It is conceivable
that the gamma-ray ``dark'' choked GRBs are even more abundant,
and would contribute more to the high energy neutrino background.

Several other neutrino mechanisms have been discussed in the 
literature. The internal shocks that power the prompt gamma-ray
emission can produce $\sim 100$ TeV neutrinos \cite{Waxman:1997ti,Wang:2008zm,Murase:2008sp},
which should be lagged behind the TeV neutrinos.
Inelastic collisions between decoupled protons and neutrons
during the acceleration of the fireball can power a multi-GeV neutrino
signal \cite{Bahcall:2000sa},
but the predicted flux level is below
the atmospheric neutrino background, and hence, difficult to detect.
The neutrinos produced in the early afterglow phase have a very high
energy ($\sim$ EeV) \cite{Waxman:2000} and are not optimized for 
detection with the present day neutrino detectors. The multi-TeV 
neutrino signal discussed here could be 
above the atmospheric background, and may be detectable with km$^3$
detectors for some nearby sources.

The main source of high energy neutrinos is the decay of charged
pions, which leads to the neutrino flux ratio at the production site
$\Phi^0_{\nu_e}:\Phi^0_{\nu_\mu}:\Phi^0_{\nu_\tau}=1:2:0$
($\Phi^0_{\nu_{\alpha}}$ corresponds to the sum of neutrino and
anti-neutrino flux for the flavor $\alpha$ at the source). 
The vacuum oscillation of these neutrinos on their way to 
Earth would make the observed ratio to $1:1:1$. 
This applies to low energy neutrinos including the TeV
neutrinos discussed in this paper.
For high energy neutrinos above $\sim$1 PeV,
muon energy is degraded before decaying to low energy
neutrinos so that high energy neutrinos will be absent.
The neutrino flux ratio at the source is modified to $0:1:0$,
which is further modified to $1:1.8:1.8$ at Earth after
vacuum oscillation is taken into account \cite{Kashti:2005qa}.

Another possibility of modifying the neutrino flavor ratio is 
the resonant conversion of neutrinos from one flavor to another due to
the medium effect. Such an effect is known to be important for solar 
neutrinos \cite{Mikheyev:1985}, and has been discussed for
hot plasma in the early universe \cite{Enqvist:1990ad}, supernova 
medium \cite{Sahu:1998jh}, the GRB fireball \cite{Sahu:2005zh,Sahu:2009iy,Sahu:2009ds}
and jet \cite{Mena:2006eq}. Here we show that for choked GRBs,
the multi-TeV neutrinos discussed by M\'esz\'aros \& Waxman
\cite{Meszaros:2001ms} could undergo resonance oscillation
in the high density core (typically He core) of the presupernova
star, if the neutrino oscillation
parameters are in the atmospheric neutrino oscillation range.
This would alter the neutrino flavor ratio escaping from the
stellar envelope, and hence, the eventual detected flavor ratio 
from Earth.

\section{Neutrino oscillation in the stellar envelope}

As a mildly relativistic jet makes its way through the stellar 
envelope, internal shocks can develop and 
can accelerate protons to energy $\sim 10^5$ GeV. These protons
would interact with the $\sim$ keV thermal X-ray photons  
to produce $\sim 5$ TeV neutrinos via the process
$p+\gamma \rightarrow \Delta^+ \rightarrow n+ \pi^+
\rightarrow n+\mu^++\nu_{\mu}\rightarrow n+
e^++\nu_{\mu}+\nu_e+{\bar \nu_{\mu}}$. 

Depending on the initial mass and metalicity, the 
presupernova star can have different compositions with
different radii. The LGRB progenitors (Type Ic SN) have lost 
the H and most of the He envelopes before explosion. They are 
too small to have an interesting neutrino oscillation signature.
The choked jet progenitors, on the other hand, can retain the 
He envelope (Type Ib SNe) and even the H envelope (Type II SNe). 
These presupernova stars are favorable for TeV neutrino production
and neutrino oscillation. 
It is believed that the putative jet is launched along the rotation
axis where the centrifugal support is the least, and is powered 
by either $\nu \bar\nu$ annihilation 
or through some electromagnetic processes.  
Without exploring the details of jet dynamics, here we 
treat the jet parametrically. We assume that the jet is developed
and the TeV neutrinos are produced at a radius $r_{\rm j} 
\ll R_{*}$,
where $R_*$ is the radius of the star.

Depending on the energy of the propagating neutrino and the nature of
the background, neutrinos can interact with the background particles
via charge current (CC) and neutral current (NC) interactions. For a
neutrino energy below $E_{\nu}\simeq M^2_W/2 m_e\simeq 10^7$ GeV,
an electron neutrino can have both CC and NC interactions with the normal
matter, whereas muon and tau neutrinos can only have NC interactions.
The effective potential of NC interactions is the same for all 
active neutrinos. Since oscillation depends only on the potential 
difference, for active-active oscillations the NC contributions cancel 
out.  So only the CC contribution to the neutrino potential,
$V=\sqrt{2} G_F N_e$, is responsible for neutrino oscillation in the 
medium, where $G_F$ is the
Fermi coupling constant and $N_e$ is the electron number density in
the medium.  For anti-neutrinos $N_e$ is replaced by $-N_e$. Thus
for the process $\nu_e\leftrightarrow\nu_{\mu,\tau}$, the neutrino
potential is $\sqrt{2} G_F N_e$, while for the process
$\nu_{\mu}\leftrightarrow\nu_{\tau}$ it vanishes.

Here we consider the simplified picture of two mixed flavor states
$\nu_e$ and $\nu_{\mu}(\nu_{\tau})$ with the vacuum mixing angle
$\theta$ and mass square difference $\Delta m^2$. In an uniform medium
the evolution of the flavor states are governed by \cite{Kim:1993,Sahu:1998jh,GonzalezGarcia:2002dz}

\be
i \frac{d}{dt}
\begin{pmatrix} {{\nu}_{e}} \\ {{\nu}_{\mu}}
\end{pmatrix}
=
\begin{pmatrix}
V-\Delta \cos 2\theta & \frac{\Delta}{2}\sin 2\theta \\
\frac{\Delta}{2}\sin 2\theta & 0\\
\end{pmatrix}
\begin{pmatrix}
\nu_{e} \\ \nu_{\mu}
\end{pmatrix},
\ee
where $\Delta=\Delta m^2/2E_{\nu}$, $V$ is the potential difference
between $V_{\nu_e}$ and $V_{\nu_{\mu}}$,
(i.e. $V=V_{\nu_e}-V_{\nu_{\mu}}$) 
and $E_{\nu}$ is the neutrino energy. The transition probability as 
a function of distance $\ell$ is given by
\be
P_{\nu_e\rightarrow {\nu_{\mu}{(\nu_\tau)}}}(\ell) = 
\frac{\Delta^2 \sin^2 2\theta}{\omega^2}\sin^2\left (\frac{\omega \ell}{2}\right
),
\label{prob}
\ee
with
\be
\omega=\left [(V-\Delta \cos 2\theta)^2+\Delta^2 \sin^2
    2\theta\right ]^{1/2}.
\ee
Once the neutrinos are produced due to pion decay at a point
$r_{\rm j} \ll R_{*}$, they will propagate away from the star where the
medium effect can be substantial. If the density of the medium is such
that the condition $\sqrt{2} G_F N_e=\Delta \cos 2\theta$ is
satisfied, then resonant conversion of neutrinos from one flavor to 
another with maximum amplitude can occur. For anti-neutrinos the 
resonance condition can never be satisfied (for
normal neutrino mass hierarchy). Although the oscillation process
${\bar \nu}_e\leftrightarrow{\bar\nu}_{\mu,\tau}$ can take place, it
will be suppressed. 

 The critical density for 
resonance is called the resonance density. For 5 TeV neutrinos, 
it reads
\be
\rho_{\rm R}=(1.32 ~{\rm g~cm^{-3}})~
\frac{\tilde {\Delta m^2}}{E_{\nu,12.7}} \cos 2\theta
\label{den}
\ee
where we have $\tilde {\Delta m^2}$ in units of eV$^2$ and 
$E_{\nu,12.7}$ in units of $10^{12.7}$ eV. The resonance length is
\be  
\ell_{\rm R}=\frac{2\pi}{\Delta \sin 2\theta}=1.24\times 10^9 ~{\rm cm}~
\left(\frac{E_{\nu,12.7}}{\tilde {\Delta m^2}}\right )
\frac{1}{\sin 2\theta}.
\label{lres}
\ee
Define the stellar radius $r_{\rm R}$ as the radius at which 
the local density is $\rho_{\rm R}$. The first condition for
resonant oscillation is $\ell_{\rm R} < r_{\rm R}$.

If the resonance region is
wide enough the transition can be total.
We can define a resonance width for which the amplitude of the
probability can be $1/2$ instead of unity. In this case the width can
be given as $\Gamma = 2 \Delta m^2 \sin 2\theta$. This corresponds
to a length scale
\be
\delta r_{\rm R}= \frac{2 \tan 2\theta}{ |\frac{1}{N_e} \frac{dN_e}{dr}|}_{\rm R}. 
\label{delr}
\ee
For $\delta r_{\rm R} > \ell_{\rm R}$, there can be enough time for 
$\nu_e$ to stay in the resonance region to convert into $\nu_{\mu}
(\nu_{\tau})$. This is the second condition for significant
resonant oscillation.

In order to evaluate both conditions, one needs to 
know the matter density profile in the stars (which determines
$r_{\rm R}$ and $dN_e/dr$). 
The density profile of a presupernova star is difficult to probe 
observationally. Numerical models predict a decreasing 
density with radius. If convective mixing is not important, 
there is a sharp decrease in density beyond the He core with radius
$r_{\rm He}\sim 10^{11}$ cm and the local density $\rho_{\rm He}
\sim 10^{-3}~{\rm g~cm^{-3}}$. If convective mixing is important,
there is no abrupt transition, and the density profile may be
roughly described in the analytical form\cite{Razzaque:2004yv,Matzner:1998mg}
\be
\rho (r)=\rho_0 \left (\frac{R_*}{r}-1 \right )^n~.
\label{rho}
\ee
The parameters $R_*$
and $\rho_0$ depend on the type of the star. For example, a blue 
supergiant (BSG) model for SN 1987A gives $R_* = 3\times 10^{12}$ cm 
and $\rho_0= 3\times 10^{-5}~{\rm g~cm^{-3}}$ \cite{Shigeyama:90}. 
In some models, the He core can extend to $10^{12}$ cm, and the
H envelope can extend to $10^{13}$ cm 
\cite{Meszaros:2001ms,Razzaque:2004yv}.

The exponent $n=3, 3/2$ correspond to the radiative
and convective envelope, respectively. In general it can
vary between $2$ and $3$ for different numerical models. 
The condition $\delta r_R > \ell_{\rm R}$ can be re-written as a 
requirement to $n$ (where $r \sim \ell_{\rm R}$ has been adopted
which is relevant for resonant oscillation)
\be
n <  2 \tan 2\theta \left (1-\frac{\ell_{\rm R}}{R_*} \right ).
\label{valn}
\ee
The resonance length $\ell_{\rm R}$ depends on neutrino oscillation
parameters and neutrino energy. 
Apparently, if $\ell_{\rm R} \geq  R_*$, the requirement to 
$n$ ($n<0$) is unphysical,
and no neutrino oscillation is expected inside these stars.
On the other hand, if $\ell_{\rm R} \ll R_*$, the constraint
$n < 2\tan 2\theta$ may be satisfied in some stars for some
oscillation parameters. 

In order to evaluate whether the neutrino oscillation
conditions are satisfied, one needs to know the neutrino
oscillation parameters in matter. Experimentally these are
inaccessible. Only oscillation parameters from the solar
and atmospheric neutrino experiments are available. 
For small neutrino mixing angles, the mixing matrix is 
almost diagonal and each flavor eigenstate nearly overlaps 
with one of the mass eigenstate. One may then associate
$\nu_e$ to $\nu_1$, $\nu_{\mu}$ to $\nu_2$, and $\nu_{\tau}$ 
to $\nu_3$. While the solar neutrino oscillation parameters
are relevant to $\nu_e \rightarrow \nu_\mu$ oscillations,
the atmospheric neutrino oscillation data correspond
mostly to $\nu_{\mu} \rightarrow \nu_{\tau}$ oscillations, and 
the corresponding neutrino parameters are $\theta_{23}$ and 
$\Delta m^2_{23}$, respectively. It is also very possible that
the physical properties of neutrinos in a medium could be 
different from their vacuum values. For example, a neutrino can 
acquire mass due to its interaction with the background 
particles even if we consider it  massless in vacuum.
Similarly the mixing properties in matter may not follow the 
vacuum pattern as measured. Nonetheless, since the oscillation
parameters of the solar and atmospheric experiments are the
best measured, in the following we test whether these parameters
may allow neutrino oscillation to happen in the progenitor stars 
of choked GRBs. We do not take it for granted 
that any of these parameters
are operating in the oscillation process $\nu_e \leftrightarrow 
\nu_{\mu,\tau}$ in the choked fireball, but just take the
only experimentally available parameters to 
test the conditions for prominent
oscillations. Similar analyses have been carried out before
to evaluate the possible oscillation effect in GRB fireballs
\cite{Sahu:2005zh,Sahu:2009iy}.

The SNO salt phase solar neutrino data, combined with the KamLand
reactor antineutrino results, constrain the neutrino oscillation
parameters in the regime $6\times 10^{-5}\, {\rm eV}^2 < \Delta m^2 <
10^{-4}\, {\rm eV}^2$ and $0.64 < \sin^2 2\theta < 0.96$ 
\cite{Ahmed:2003kj}, with the best fit parameters
$\Delta m^2\sim 7.1\times 10^{-5}\, {\rm eV}^2$ and $\sin^2 2\theta \sim
0.69$ with $99\%$ confidence level. The best fit values give
$\rho_{\rm R,SNO}\simeq 5.2\times 10^{-5}\, {\rm g~cm}^{-3} 
E_{\nu, 12.7}^{-1}$ 
and $\ell_{\rm R,SNO}\simeq 2.1\times 10^{13} {\rm cm} E_{\nu, 12.7}$. 
We can see that $\ell_{\rm R}$ is larger than $R_*$ of a typical BSG,
suggesting that resonant oscillation would not occur for these neutrino
oscillation parameters.

On the other hand, the atmospheric
neutrino oscillation parameters reported by the SK Collaboration are
in the range $1.9\times 10^{-3}\, {\rm eV}^2 < \Delta m^2 < 3.0\times
10^{-3}\, {\rm eV}^2$ and $0.9 \le \sin^2 2\theta \le 1.0$ at a $90\%$
confidence level \cite{Ashie:2004mr},  which 
corresponds to the oscillations of mostly muon neutrinos to 
tau neutrinos. If we assume that these parameters apply to
$\nu_e \rightarrow \nu_{\mu,\tau}$ oscillations in matter, 
we can get the following constraint.
We consider the good fit point 
$\Delta m^2 \sim 2.5\times 10^{-3}\, {\rm eV}^2$
and $\sin^2 2\theta \sim 0.9$, and get
$\rho_{\rm R,SK}\simeq 1.0\times 10^{-3}~{\rm g\,cm}^{-3}~E_{\nu,
12.7}^{-1}$ and
$\ell_{\rm R,SK}\simeq 5.2\times 10^{11}~{\rm cm}~E_{\nu, 12.7}$.   
For the nominal BSG model discussed in this paper, the stellar 
radius at which the density is $\rho_{\rm R,SK}$ is $r_{\rm R}
\simeq 7.1\times 10^{11}$ cm for $n=3$ and $E_{\nu,12.7}=1$. 
We can see that the condition $\ell_{\rm R} < r_{\rm R}$ is 
satisfied for the typical neutrino energy $E_{\nu} = 5$ TeV.
By using the value of $\tan 2\theta$ from the SK neutrino data and
$\ell_{\rm R}=\ell_{\rm R,SK}$, we obtain from Eq.~(\ref{valn}) 
$n < 4.96$. Known stellar models have $n$ between 2 and 3. This 
suggests that the second condition is also fully satisfied.
We conclude that resonant oscillation of multi-TeV neutrinos
can occur within a nominal BSG progenitor for the neutrino
oscillation parameters inferred by the atmospheric neutrino
data. A similar analysis suggests that the same conclusion
applies to other BSG progenitors or He stars with
extended envelopes (with $R_*$ up to $10^{12}$ cm), but does
not apply to typical He stars (with $R_* = 10^{11}$ cm), or 
other more compact stars.
Since GRB observations favor associations with Type Ic SNe
(for which the He envelope is mostly stripped off), GRBs
are not preferred sources for TeV neutrino oscillation.
Instead, we identify choked GRBs, especially those with
a heavy envelope, as interesting sources for resonant
neutrino oscillation.

For a full oscillation, $\nu_e$ can oscillate to $\nu_{\mu}$ and
to $\nu_{\tau}$ with equal probability but $\nu_{\mu}$ can oscillate
only to $\nu_e$.  On average we can have $1/3$ survival probability
for $\nu_e$, $\nu_{\mu}$ and $\nu_{\tau}$ for each $\nu_e$
oscillation, but have 1/2 survival probability for $\nu_e$,
$\nu_{\mu}$ for the $\nu_{\mu}\leftrightarrow \nu_e$ resonant
oscillation. The $p\gamma$ process also produce 
$\bar \nu_{\mu}$ which does not resonantly oscillate. 
Putting together, on the surface of the presupernova
star the survival probability of each flavor (both neutrino and
anti-neutrino) is in the ratio of 
$(\frac{1}{3}+\frac{1}{2}):(\frac{1}{3}+\frac{1}{2}+1):\frac{1}{3}=
5:11:2$. Above $r_R$ (mostly in the H envelope), the density is
much lower than the resonance density, and no further reconversion 
of neutrinos can take place. So the ratio $5:11:2$ is final neutrino 
flavor ratio escaping from the star.
This ratio is notably different from 
the nominal $1:2:0$ ratio for multi-TeV neutrinos without 
considering the resonant oscillation effect.
 
\section{Neutrino oscillation in vacuum}

Since these choked GRB neutrino sources are typically at
large distances, the TeV neutrinos escaping from the star
would undergo vacuum oscillation on their way to Earth. 
The neutrino flux for a particular flavor $\alpha$ on Earth is given by
\be
\Phi_{\nu_{\alpha}}=\sum_{\beta} P_{\alpha\beta} \Phi^s_{\nu_{\beta}},
\label{fluxE}
\ee
where $\Phi^s_{\nu_{\beta}}$ signifies the flux of $\nu_{\beta}$ at 
the surface of the He envelope after resonant oscillation, and 
$P_{\alpha\beta}$ corresponds to the oscillation probability from 
${\nu_{\alpha}}$ to $\nu_{\beta}$ in vacuum.  
For the matter effect on resonant oscillation in the stellar envelope,
we have applied a two-flavor neutrino analysis. 
This is because no $\nu_{\tau}$'s are generated in the $p\gamma$
process on one hand, and the two-flavor neutrino analysis is 
simpler on the other hand, as it depends only on one mass square 
difference. The limited size of the oscillation baseline (the 
stellar envelope) also makes the three-flavor oscillation effect 
unimportant. Such a two-flavor neutrino 
analysis has been applied in most previous resonant oscillation
discussions \cite{Sahu:2005zh,GonzalezGarcia:2002dz}.

When discussing the vacuum oscillation effect along a long base line
from the source to Earth, one needs however fully take into account 
the three-flavor neutrino oscillation effect. This is demanded by 
the combined analyses of both the solar and the atmospheric
neutrino anomalies.  
For the best fit to
the SNO data from the large mixing angle (LMA) solution one can take
the mixing angles $\theta_{12}=34^{\circ}\pm 2.5^{\circ}$,
$\theta_{23}=45^{\circ}\pm 6^{\circ}$, $\theta_{13}=0^{\circ}\pm
8^{\circ}$ and the Dirac phase
$\delta=0$ \cite{Kashti:2005qa,Strumia:2005tc}. This gives
$P_{ee}\simeq 0.57$, $P_{e\mu}=P_{e\tau}\simeq 0.215$ and
$P_{\mu\mu}=P_{\mu\tau}=P_{\tau\tau}\simeq 0.393$. Inserting these
probabilities in Eq.~(\ref{fluxE}) and considering the error
of $\theta_{12}$, the flux ratio at Earth is 
$1:(1.095\pm 0.012):(1.095\pm 0.012)$.
On the other hand, if we consider the maximal mixing among the
$\nu_{\mu}$ and $\nu_{\tau}$ in vacuum, then the $\nu_e$
oscillation to $\nu_{\mu}$ and $\nu_{\tau}$ is largely suppressed.
One can then have $P_{ee}\simeq 1$, $P_{\mu\mu}=P_{\tau\tau}=
P_{\mu\tau}=1/2$, and all other transition probabilities are
negligible. Using these oscillation probabilities, we obtain the
flux ratio at Earth as 1:1.3:1.3.

\section{Discussion}

GRBs have a wide redshift distribution (from $z=0.0085$ to $z=8.2$).
Observations suggest that the nearby low-luminosity GRBs have a
local event rate $\sim 100$ time higher than that of high-luminosity
GRBs \cite{Liang:2007}. If GRB jets become progressively successful
in progressively rarer progenitors, it is conceivable that there
could be even more choked GRBs in the nearby universe. Although
the gamma-ray luminosity becomes progressively smaller as the
envelope becomes progressively heavier, the TeV neutrino luminosity 
may not decrease, and could even follow
an opposite trend. Assuming that the choked GRB progenitor has
a local event rate similar to that of LL-GRBs, i.e. 
${\cal R} \sim 200 ~{\rm Gpc^{-3}~yr^{-1}}$, one would expect 
$\sim 14$ neutrino bursts (without gamma-ray counterpart) 
per year all sky at $z<0.1$. If the TeV neutrino 
luminosity of these events are similar to those of successful
GRBs \cite{Meszaros:2001ms,Razzaque:2004yv}, then each event would
have 100s of TeV neutrinos detected by a km$^3$ detector such as 
IceCube. Such a possibility is already ruled out by the current
upper limits placed by the IceCube observations. This suggests that
either there are not that many nearby choked GRBs, or that the
choked GRBs are not as neutrino-luminous as predicted
\cite{Meszaros:2001ms}. Going to the conservative extreme, i.e. if 
the neutrino luminosity is correlated with gamma-ray luminosity, 
then the detected event rate for these nearby neutrino burst
sources would be 0.001-0.01 neutrinos per event in a km$^3$
detector. This is essentially impossible for IceCube to detect
individual sources. The real detected neutrino event rate may be 
between these two extreme values. IceCube or a similar 
detector would be able to detect these neutrino bursts or to 
place even more stringent upper limits in the near future.

If the nearby neutrino bursts are bright enough, 
the deviation of the 
observed neutrino flavor ratio from 1:1:1 may be tested by 
IceCube or similar detectors. The flavor ratios
can be in principle deduced from the relative rates of showers,
muon tracks, and the unique tau lepton induced signals 
\cite{Beacom:2003nh}.
The possibility of detecting a tau signal by IceCube is low,
especially in the multi-TeV energy range. On the other hand,
IceCube can distinguish between  shower-like events and the 
$\mu$-track events, although it is hard to identify $\nu_e$ 
and $\nu_{\tau}$ through their electromagnetic and hadronic 
showers. Nonetheless, assuming a flavor-independent neutrino
spectrum and $\nu_{\mu}-\nu_{\tau}$ symmetry (as is the case
of our two predicted ratios), $\nu_e$ fraction may be extracted 
from the measured Muon to shower ratio 
\cite{Beacom:2003nh}. 
The 10$\%$ difference in the flavor ratio reduces the $\nu_e$ fraction
from 1/3 to 0.313 (for flavor ratio 1:1.095:1.095). This corresponds to
a slight increase of Muon to shower ratio. With the uncertainty
($20\%$) for the nominal diffuse flux 
($E_{\nu_\mu}^2 d N_{\nu_\mu}/dE_{\nu_\mu}=10^{-7}~{\rm GeV~cm^{-2}
~s^{-1}}$ for one year) adopted in \cite{Beacom:2003nh}, the
small change in Muon to shower ratio may not be differentiated.
If nearby neutrino bursts are bright enough, the flux would be
increased and the uncertainty reduced significantly.
This would make a better case for 
detecting the flavor ratio change.
For the 1:1.3:1.3 ratio, the $\nu_e$ fraction is reduced from 1/3
to 0.28, making the Muon to shower ratio as high as $\sim$ 3.5
(as compared to $\sim 3$ for 1:1:1). The effect may be detectable
for the putative bright neutrino burst events discussed above,
even if they may be very rare.

Since the parameters ($\ell_{\rm R}$, $\rho_{\rm R}$, $r_{\rm R}$,
and $\delta r_{\rm R}$) all depend on $E_\nu$, we expect the
flavor ratio would also depend on neutrino energy. This aspect
has been extensive discussed in \cite{Razzaque:2009kq}.

This work is partially supported by DGAPA-UNAM (Mexico) Project 
No. IN101409 and Conacyt project No. 103520 (SS) and by NASA NNX09AO94G and NSF AST-0908362 (BZ).

\end{document}